\documentclass[PP]{AEA}

\usepackage{natbib}

\draftSpacing{1.5}

\newcommand{\obs}{\mathrm{obs}}
\newcommand{\independent}{\perp\!\!\!\perp}

\newcommand{\mme}{\mathbf{E}}
\newcommand{\mmv}{\mathbf{V}}
\newcommand{\mmav}{\mathbf{AV}}
\newcommand{\prr}{{\rm pr}}

\def\monthname{\ifcase\month\or
  January\or February\or March\or April\or May\or June\or July\or
  August\or September\or October\or November\or December\fi}
\numberwithin{equation}{section}

\begin{document}

\title{Estimating Average Treatment Effects: Supplementary Analyses and Remaining Challenges}
\shortTitle{Average Treatment Effects}
\author{Susan Athey, Guido Imbens, Thai Pham, and Stefan Wager\thanks{Athey: Graduate School of Business, Stanford University, athey@stanford.edu. 
Imbens: Graduate School of Business, Stanford University, imbens@stanford.edu. 
Pham: Graduate School of Business, Stanford University, thaipham@stanford.edu.
Wager: Department of Statistics, Columbia University, and  Graduate School of Business, Stanford University, swager@stanford.edu.
We are grateful for discussions with Jasjeet Sekhon and comments by Panos Toulis}}\pubMonth{Month}
\date{\today}
\pubYear{Year}
\pubVolume{Vol}
\pubIssue{Issue}
\maketitle

\section{Introduction}

There is a large literature in econometrics and statistics on semiparametric estimation of  average treatment effects under the assumption of unconfounded treatment assignment. Recently this literature has focused on the setting with many covariates, where regularization of some kind is required. In this article we discuss some of the lessons from the earlier literature and their relevance for the many covariate setting, and propose some supplementary analyses to assess the credibility of the results.

\section{The Set Up}

We are interested in estimating an average treatment effect in a setting with a binary treatment. We use the potential outcome or Rubin Causal Model set up (\citet{rubin1974estimating, holland1986statistics, imbens2015causal}). Each unit in a large population is characterized by a pair of potential outcomes $(Y_i(0),Y_i(1))$, with the  estimand equal to the average causal effect:
\[ \tau=\mme[Y_i(1)-Y_i(0)],\]
or the average effect for the treated, $\tau_t=\mme[Y_i(1)-Y_i(0)|W_i=1]$.
The treatment assignment for unit $i$ is $W_i\in\{0,1\}$. For each unit in a random sample from the population we observe the treatment received and the realized outcome,
\[ Y_i^\obs=
\left\{\begin{array}{ll}
Y_i(0)\hskip1cm & {\rm if}\  W_i=0,\\
Y_i(1)\hskip1cm & {\rm if}\  W_i=1,
\end{array}\right.\]
and pretreatment variables or features $X_i$.
To identify $\tau$ we assume unconfoundedness  (\citet{rosenbaum1983central}) 
\[ W_i\ \independent\ \Bigl(Y_i(0),Y_i(1)\Bigr)\ \Bigl|\ X_i,\]
and overlap of the covariate distributions,
\[e(x)\in(0,1),\] where
the propensity score (\citet{rosenbaum1983central}) is $e(x)=\prr(W_i=1|X_i=x)$. Define the marginal treatment probability $p=\mme[W_i]$, the conditional means of the potential outcomes, $\mu(w,x)=\mme[Y_i(w)|X_i=x]$, the marginal means, $\mu_w=\mme[Y_i(w)]$, and the conditional variances $\sigma^2(w,x)=\mmv(Y_i(w)|X_i=x)$.
The efficient score  for $\tau$, which plays a key role in the discussion,  is
\[ \phi(y,w,x;\tau,\mu(\cdot,\cdot),e(\cdot))=
w \frac{y-\mu(1,x)}{e(x)}-\]
\[
(1-w) \frac{y-\mu(0,x)}{1-e(x)}
+\mu(1,x)-\mu(0,x)-\tau,\]
(\citet{hahn1998role}) and the implied  semiparametric variance bound
 is
\[ \mmav=\mme\left[\phi(Y^\obs_i,W_i,X_i;\tau,\mu(\cdot,\cdot),e(\cdot))^2\right].\]
For the average effect for the treated, $\tau_t$, the efficient score function is
\[ \phi'(y,w,x;\tau_t,\mu(\cdot,\cdot),e(\cdot))=
\frac{w}{p} (y-\mu(0,x)-\tau)\]
\[+
\frac{(1-w)e(x)}{p(1-e(x))}(y-\mu(0,x)).\]
A  wide range  estimators for  $\tau$  have been proposed in this setting,  (see for a review \citet{imbenswooldridge}).
Some of the proposed estimators rely on   matching  (\citet{abadie2006}). Others rely on different characterizations of the  average treatment effect, using the propensity score,  (\citet{hirr}),
\[ \tau=\mme\left[\frac{Y_i^\obs\cdot W_i}{e(X_i)}-\frac{Y_i^\obs\cdot (1-W_i)}{1-e(X_i)}\right],\]
the conditional expectation of the outcome,
\[ \tau=\mme\Bigl[\mu(1,X_i)-\mu(0,X_i)\Bigr],\]
 (\citet{hahn1998role}),
or the efficient score representation
\[ \tau=\mme\Biggl[ 
W_i \frac{Y^\obs_i-\mu(1,X_i)}{e(X_i)}+\]
\[ 
(1-W_i)\frac{Y^\obs_i-\mu(0,X_i)}{1-e(X_i)}
+\mu(1,X_i)-\mu(0,X_i)\Biggr]\]
(\citet{van2000asymptotic, van2006targeted, chernozhukov2016double}). Corresponding estimators exist for the average effect for the treated.

Because the unconfoundedness assumption imposes no restrictions on the joint distribution of the observed variables $(Y^\obs_i,W_i,X_i)$,  it follows by the general results for semiparametric estimators in \citet{newey1994asymptotic} that all three approaches, substituting suitable (sometimes undersmoothed) nonparametric estimators of the propensity score and/ or the conditional expectations of the potential outcomes, and replacing the expectations by averages, reach the semiparametric efficiency bound. 

\section{Four Issues}

First we wish to raise four issues that have come up in the fixed-number-of-covariate case, and which are even more relevant in the many covariate setting.

\subsection{Double Robustness}

A consistent finding from the observational study literature with a fixed number of pretreatment variables is that the best estimators in practice involve both estimation of the conditional expectations of the potential  outcomes and estimation of the propensity score, rendering them less sensitive to estimation error in either. (Although this  appears to be less important in the case of a randomized experiment, where simply estimating the conditional expectation of the outcome automatically leads to robustness, e.g., \citet{wager2016high} because the propensity score is constant and therefore always correctly specified.)
An important notion in the observational study literature, formalizing this idea, is that of so called ``doubly robust'' estimators (\citet{robins1, robins2, scharfstein1999adjusting}) that  rely for consistency only on consistent estimation of either the propensity score or the conditional outcome expectations, but do not require consistent estimation of both. See also, \citet{kang2007demystifying} for a critical perspective on these ideas.
As a simple example to develop intuition for this,
consider the standard omitted variable bias formule when estimating a regression function
\[ Y^\obs_i= W_i\tau+ X_i\beta+\varepsilon_i.\]
Omitting $X_i$ from this regression leads to a bias in the least squares estimator for $\tau$ if, first, the included regressor $W_i$ and the omitted regressor $X_i$ are correlated, and second, the omitted regressor has a non-zero coefficient. In this setting weighting by the inverse of, or conditioning on,  the propensity score removes the correlation between $W_i$ and $X_i$. Therefore it eliminates the sensitivity to the parametric form in which $X_i$ is included, without introducing bias if the weights are misspecified but the regression function is correct.

Here we view estimators as at least in the spirit of doubly robust estimation if they attempt to adjust directly for the association between the treatment indicator and the covariates, through balancing,  weighting, or otherwise, and adjust directly for the association between the potential outcomes and the covariates.
There are multiple ways of obtaining such estimators. One can do so by subclassification on the propensity score in combination with regression within the subclasses, or weighting in combination with regression. For example, suppose we parametrize the conditional means as $\mu(w,x)=w\tau+x'\beta$, and the propensity score as $e(x)=1/(1+\exp(x'\gamma))$, and  estimate the regression by weighted linear regression with weights equal to $W_i/\sqrt{e(X_i;\hat\gamma)}+(1-W_i)/\sqrt{(1-e(X_i;\hat\gamma))}$, then the estimator for $\tau$  is consistent if either the propensity score or the conditional expectations of the potential outcomes are correctly specified.
Similarly, using the efficient score, if we estimate the average treatment effect by solving
\[\frac{1}{N}\sum_{i=1}^N \phi\Bigl(Y^\obs_i,W_i,X_i;\tau,\hat\mu(\cdot,\cdot),\hat e(\cdot)\Bigr)=0,\]
as a function of $\tau$ given estimators $\hat\mu(\cdot,\cdot)$ and $\hat e(\cdot)$,
then as long as either the estimator for either $\mu(w,x)$ or  $e(x)$ is consistent, the resulting estimator for $\tau$ is consistent.

If we use general nonparametric estimators for $\mu(\cdot),\cdot)$ and $e(\cdot)$, this last estimator also has the property that the  estimator for the finite dimensional component $\tau$ is asymptotically uncorrelated with the estimator for the nonparametric components $\mu(w,x)$ and $e(x)$. This orthogonality property  (\citet{chernozhukov2016double})  follows  from the representation of the estimator in terms of the efficient score. 
Note that the properties are distinct: not all estimators that have the orthogonality property are doubly robust.

\subsection{Modifying the Estimand}

A second issue is the choice of estimand. Much of the literature has focused on the average treatment effect $\mme[Y_i(1)-Y_i(0)]$, or the average effect for the treated. A practical concern is that these estimands may be difficult to estimate precisely if the propensity score is close to zero  for a substantial fraction of the population.
This is a particular concern in settings with many covariates because regularization based on prediction criteria may downplay biases that are present in estimation of $\mu(w,x)$ in parts of the $(w,x)$ space with few observations, even if those values are important for the estimation of the average treatment effect. 
 In that case one may wish to focus on a weighted average effect of the treatment. One can do so by trimming or weighting. 
\citet{ crump2006moving, crump2009dealing} and \citet{li2014balancing} 
suggest estimating
\[\tau_{\omega(\cdot)}=\frac{\mme\left[\omega(X_i)\cdot\left(Y_i(1)-Y_i(0)\right)\right]}{\mme\left[\omega(X_i)\right]},\]
for $\omega(x)=e(x)(1-e(x))$ or $\omega(x)={\bf 1}_{\alpha<e(x)<1-\alpha}$. 
The semiparametric efficiency bound for $\tau_{\omega(\cdot)}$ is  (\citet{hirr})
\[ \mmav =\frac{1}{\mme[\omega(X_i)^2]}\mme\biggl[
\frac{\omega(X_i)^2\sigma^2(1,X_i)}{e(X_i)}\]
\[+
\frac{\omega(X_i)^2\sigma^2(0,X_i)}{1-e(X_i)}\]
\[ +\omega(X_i)^2\left(\mu(1,X_i)-\mu(0,X_i)-\tau_{\omega(\cdot)}\right)^2
\biggr],\]
which can be an order of magnitude smaller than the asymptotic variance bound for $\tau$ itself.

In settings with limited or no heterogeneity in the treatment effects as a function of the covariates, these weights are particularly helpful and the weights $\omega(x)=e(x)(1-e(x))$ lead to efficient estimators for $\tau$ in that case.
The arguments in 
\citet{ crump2006moving, crump2009dealing} and \citet{li2014balancing} show that one may wish to impose a constant treatment effect in estimation even if substantively one does not find that assumption credible.

\subsection{Weighting versus Balancing}
\label{balancing}

Although weighting by the inverse of the treatment assignment balances pretreatment variables in expectation, it does not do so in finite samples. Recently there have been a number of estimators proposed that focus directly on balancing the pretreatment variables, bypassing estimation of the propensity score
(\citet{hainmueller, zubizarreta2015stable, graham1, graham2, atheyimbenswager}).
Specifically, given a set of pretreatment variables $X_i$, one can look for a set of weights $\lambda_i$ such that
\[ \frac{1}{N_t}\sum_{i=1}^N \lambda_i\cdot W_i \cdot X_i\approx 
\frac{1}{N_c}\sum_{i=1}^N \lambda_i\cdot (1-W_i) \cdot X_i,\]
where $N_c$ and $N_t$ are the number of control and treated units respectively. The advantage of such weights is that they eliminate any biases associated with linear and additive effects in the pretreatment variables in the estimator
\[ \hat\tau=\frac{\sum_{i=1}^N \lambda_i W_i Y^\obs_i}{\sum_{i=1}^N \lambda_i W_i}
-
\frac{\sum_{i=1}^N \lambda_i (1-W_i) Y^\obs_i}{\sum_{i=1}^N \lambda_i (1-W_i)},\]
whereas using the propensity score weights $\lambda_i=W_i/e(X_i)+(1-W_i)/(1-e(X_i))$ does so only in expectation.

\subsection{Sensitivity}

Consider the simple difference in average outcomes by treatment status as an estimator for the average treatment effect.
The bias in this estimator arises from the presence of pretreatment variables that are associated with both the treatment and the potential outcomes. Pretreatment variables that are associated solely with the treatment, or solely with the potential outcomes may make it difficult to estimate the propensity score or the conditional expectations of the potential outcomes, but such variables do not compromise the estimates of the average treatment effects. As a result it is not so much sparsity of the propensity score or sparsity of the conditional expectations, but sparsity of the product of the respective coefficients that matter. A summary measure of this association is the characterization of the bias as an expected value,
\[ {\rm B}=\left(\mme[Y^\obs_i|W_i=1]-\mme[Y^\obs_i|W_i=0]\right)-\tau\]
\[\hskip1cm =\frac{1}{p(1-p)})\mme\left[b(X_i)\right],\]
where the bias function $b(\cdot)$ is
\[ b(x)=( e(x)-p)\]
\[\times (p(\mu(0,x)-\mu_0)+(1-p)(\mu(1,x)-\mu_1)
.\]
Hence  the bias is proportional to the covariance of the propensity score and a weighted average of the conditional expectations of the potential outcomes,
\[{\rm Cov}\Bigl(e(X_i),p\mu(0,X_i)+(1-p)\mu(1,X_i)\Bigr).\]
The bias function at $x$ measures the contribution to the overall bias $B$, coming from units with $X_i=x$. It is flat in a randomized experiment, or in cases where the pretreatment variables are not associated with the outcome.
As another special case, consider a setting where all the pretreatment variables are uncorrelated and have mean zero and unit variance. If $e(x)=x'\gamma$, and $\mu(w,x)=\tau w+x'\beta$, then $b(x)=\beta'xx'\gamma/(p(1-p))$, so that $B=\beta'\gamma/(p(1-p))$, depending only on the product of the coefficients in the outcome equation and the propensity score.

Settings where the bias ${\rm B}$ is large relative to the difference in average outcomes by treatment effects, or $b(\cdot)$ is very variable, are particularly challenging for estimating $\tau$. In our calculations below we report summary statistics of $\hat b(X_i)$, scaled by the standard deviation of the outcome.

\section{Three Estimators}

Here we briefly discuss three of the most promising estimators that have been proposed for the case with many pretreatment variables. All three address biases from the association between pretreatment variables and potential outcomes and between pretreatment variables and treatment assignment. There are  other estimators using machine learning methods that focus only on one of these associations, for example inverse propensity score weighting estimators that estimate the propensity score using machine learning methods (\citet{mccaffrey2004propensity}), but we do not expect those to perform well. The first two estimators we discuss assume linearity of the conditional expectation of the potential outcomes in the, potentially many, covariates. How sensitive the results are in practice to this linearity assumption in settings with many covariates, where some of the covariates may be functions of underlying variables, remains to be seen.

\subsection{The Double Selection Estimator (DSE)}

\citet{belloni2013program} propose using LASSO (\citet{tibshirani1996regression}) as a covariate selection method. They do so first to select pretreatment variables that are important for explaining the outcome, and then to select pretreatment variables that are important for explaning the treatment assignment. They then combine the two sets of pretreatment variables and estimate a regression of the outcome on the treatment indicator and the union of the selected pretreatment variables.

\subsection{The Approximate Residual Balancing Estimator (ARBE)}

\citet{atheyimbenswager} suggest using elastic net (\citet{zou2005regularization}) or LASSO  (\citet{tibshirani1996regression}) to estimate the conditional outcome expectation, and then using an approximate balancing approach in the spirit of 
\citet{zubizarreta2015stable} as discussed in Section \ref{balancing} to further remove bias arising from remaining imbalances in the pretreatment variables.

\subsection{The  Doubly Robust Estimator (DRE) and the Double Machine Learning Estimator (DMLE)}

In the general discussion of semiparametric estimation \citet{van2000asymptotic} suggest estimating the finite dimensional component as the average of the influence function, with the infinite dimensional components estimated nonparametrically, leading to a doubly robust estimator in the spirit of \citet{robins1, robins2, scharfstein1999adjusting}. In the specific context of estimation of average treatment effects \citet{van2006targeted} propose this estimator as a special case of the targeted maximum likelihood approach, suggesting various machine learning methods for estimation of the conditional outcome expectation and the propensity score. \citet{chernozhukov2016double}, in the context of much more general estimation problems,  propose a closely related estimator focusing on the orthogonality properties arising from the use of the efficient score. In the  \citet{chernozhukov2016double} approach the sample is partitioned into $K$ subsamples, with the nonparametric component estimated on one subsample, and the parameter of interest estimated as the average of the influence function over the remainder of the sample. This is repeated $K$ times, and the estimators for the parameter of interest averaged to obtain the final estimator, thereby further improving the properties in settings with many covariates. We report both the simple version of the DRE and the averaged version DMLE.

\section{Outstanding Challenges and Practical Recommendations}

Here we present some practical recommendations for researchers estimating treatment effects, and discuss some of the remaining challenges for the theoretical researchers.

\subsection{Recommendations}

The main recommendation is to report analyses beyond the point estimates and the associated standard errors. Supporting analyses should be presented to convey to the reader that the estimates are credible (\citet{athey2016state}). By credible we do not mean whether the unconfoundedness property holds, but whether the estimates effectively adjust for differences in the covariates. Here are four specific recommendations to do so.

\begin{enumerate}
\item{\bf (Robustness)} Do not rely on a single estimation method. Many of the methods have attractive properties under slightly different sets of regularity conditions but rely on the same fundamental  set of identifying assumptions. These regularity conditions are difficult to assess in practice. Therefore, if the substantive  results are not robust to the specific choice of estimator, it is unlikely that the results are credible.
\item {\bf (Overlap)} Assess concerns with overlap by comparing the variance bound for $\tau$ and $\tau_{\omega(\cdot)}$ for
a choice of $\omega(\cdot)$ that de-emphasizes parts of the covariate space with limited overlap.
 If there is a substantial efficiency difference between the  $\tau$ and $\tau_{\omega(\cdot)}$,  report results for both.
\item{\bf (Specification Sensitivity)} Split the sample based on median values of each of the covariates in turn, estimate the parameter of interest on both subsamples and average the estimates to assess sensitivity to the model specification (e.g., \citet{athey2015measure}).
\item{\bf (Half Sample Bias Estimates)} Report half-sample estimates of the bias of the estimator, calculated as the estimator minus the average of estimates based on half samples, created by repeatedly randomly splitting the original sample into two equal-sized subsamples (\citet{efron1994introduction}). Asymptotic results rely on bias components of the asymptotic distribution vanishing. These estimates may shed light on the validity of such approximations. 
For example, it could reveal sensitivity to the choice of regularization parameter.
\end{enumerate}

\subsection{Some Illustrations}

Here we illustrate these recommendations with  the \citet{connors1996effectiveness} heart catherization data, with 72 covariates. 
In a working paper version we provide two additional illustrations based on the Lalonde data.
We report six estimators, the simple difference in average outcomes by treatment status, the OLS estimator with all covariates, the DS estimator (\citet{belloni2013program}), the ARB estimator (\citet{atheyimbenswager}), and the DR and DML estimators (\citet{van2006targeted, chernozhukov2016double}). 
In addition to the point estimates, we report simple bootstrap standard errors,  the scaled bootstrap bias (SBB, calculated as the average difference between the estimates, based on equal size sample splits, and the overall estimate, scaled by the bootstrap standard error. 
In addition we report average of the estimator based on sample splits, one for each covariate, where we split the sample by the median value of each covariate in turn. Given the splits we calculate the estimator for each of the two subsamples, and then average those.
See \citet{athey2015measure} for details.
We also report summary statistics of $\hat b(X_i)$, the average, the median and the 0.025, 0.25, 0.75  and 0.975 quantiles, based on random forest methods.
We also present a histogram of $\hat b(x)$fs.


For the \citet{connors1996effectiveness} data the methods do vary substantially, with the four estimators (ignoring the naive difference in means and the ols estimator) ranging from 0.038 to 0.062. This range is substantial compare to the difference relative to the naive estimator of 0.074, and relative to the standard error.
Trimming does not reduce this range substantially. The scaled bootstrap bias is as large as 29\% of the standard error, so coverage of confidence intervals may not be  close to nominal. Splitting systematically on the 70 covariates generates substantial variation in the estimates, with the standard deviation of the estimates (around 0.10) of the same order of magnitude as the standard errors of the original estimates (around 0.14).  The tentative conclusion is that under unconfoundedness the average effect is likely to be positive, but with a range substantially wider than that captured by the confidence intervals based on any of the estimators.

 \begin{table}[ht]
 \caption{\sc An Illustration Based on the \citet{connors1996effectiveness} Heart Catherization Data}
 \begin{center} 
\begin{tabular}{lcccccccc}
Metric    &           &             & trimmed &            & \multicolumn{2}{c}{Cov Split} \\ 
  	     &   ATT  & (s.e.)    & ATT        &  SBB   & mean  & std      \\ \hline
Naive & 0.074 & 0.014 & 0.038 & -0.002 & 0.073 & 0.011 \\ 
  OLS & 0.064 & 0.014 & 0.056 & 0.704 & 0.073 & 0.011 \\
  DSE & 0.062 & 0.014 & 0.057 & -0.213 & 0.061 & 0.007 \\ 
  ARBE & 0.061 & 0.015 & 0.050 & -0.157 & 0.061 & 0.007 \\
  DRE & 0.038 & 0.012 & 0.039 & 0.084 & 0.039 & 0.006 \\  
  DMLE & 0.037 & 0.014 & 0.036 & 0.341 & 0.042 & 0.007 \\ 
&&&\multicolumn{5}{c}{Quantiles}\\
 && mean  & 0.025& 0.250 & 0.500 & 0.750  & .975& \\
\multicolumn{2}{c}{$\hat b(X_i)/{\rm std}(Y_i)$}  & 0.07 & -1.29 & -0.54 & 0.25 & 0.58 & 1.29\\
 \end{tabular}

  \end{center}
 \label{tabel1}
\end{table}

\subsection{Challenges}

There are now more credible methods available for estimating average treatment effects under unconfoundedness with many covariates than there used to be, but there remain challenges in making these methods useful to practitioners. Here are some of the challenges remaining.

\begin{enumerate}
\item{\bf (Choice of Regularization)}  The regularization methods used continue to be based on optimal prediction for the infinitely dimensional components of the influence function.
Although in some cases this may be optimal in large samples, e.g., \citet{wager2016high}, 
 in many cases these methods do not focus on the ultimate object of interest, the average treatment effect, and the implication that not all errors in estimating the unknown functions matter equally.  See for some discussion of this issue \citet{athey2015recursive}. 
\item{\bf (Choice of Prediction Methods)}  
The leading estimators allow for the use of many different prediction methods of the infinitely dimensional components, without guidance for practioners how to choose among these methods in practice.
\item{\bf (Supporting Analyses)}  There is more work needed on supporting analyses that are intended  to provide evidence that in a particular data analysis the answer is credible.
\end{enumerate}

\bibliographystyle{plainnat}
\bibliography{references}

\end{document}